\documentclass[twocolumn,showpacs]{revtex4}
\usepackage{psfig}

\begin{document}

\newlength{\figureheight}
\setlength{\figureheight}{7.cm}
\newlength{\figurewidth}
\setlength{\figurewidth}{8.5cm}
\newlength{\figureheightlong}
\setlength{\figureheightlong}{9.cm}
\newlength{\figureheighttwo}
\setlength{\figureheighttwo}{12.cm}
\newlength{\figurespace}
\setlength{\figurespace}{-4mm}

\title{Phenomenological analysis of $K^+$-meson production in
  proton-nucleus collisions} 
\author{M. B\"uscher$^1$,  B.L. Ioffe$^2$, V. Koptev$^3$,
  M. Nekipelov$^{1,3}$, A. Sibirtsev$^1$, K. Sistemich$^1$, J. Speth$^1$,
  H. Str\"oher$^1$}
\affiliation{$^1$Institut f\"ur Kernphysik, Forschungszentrum J\"ulich,
  52425 J\"ulich, Germany\\
  $^2$Institute of Theoretical and Experimental Physics,
  117259 Moscow, Russia \\
  $^3$High Energy Physics Department, Petersburg Nuclear
  Physics Institute, 188350 Gatchina, Russia}
\date{\today}

\begin{abstract}
  {We investigate the experimental data, total and differential cross
    sections, on the production of $K^+$-mesons in $pA$ interactions
    at projectile energies between $T_p=0.8$ and 2.9 GeV, covering the
    transition across the free nucleon-nucleon threshold at 1.58 GeV.
    No clear evidence for the expected change of the dominant reaction
    mechanism from two-step to direct kaon production is found.  It is
    suggested that further data, in particular at forward angles, are
    taken in order to clarify the situation. It also is shown that,
    independent of the beam energy and emission angle, the invariant
    $K^+$-production cross sections show an overall exponential
    scaling behaviour with the squared four-momentum transfer between
    the beam proton and the produced $K^+$-meson for
    $t{<}{-}0.05$~GeV$^2$. The most recent data from COSY-J\"ulich,
    differential cross sections measured for $t{>}0$~GeV$^2$, show a
    strongly different $t$ dependence. Further data at forward angles
    and different beam energies are needed in order to exploit this
    region of kinematically extreme conditions.}
\end{abstract}

\pacs{29.40.-n; 25.40.-h\\
Keywords:  $K^+$-mesons; Meson production; Medium modifications
}
\maketitle

\section{Introduction}
The study of sub- and near-threshold proton-induced production of
$K^+$-mesons in nuclei has received considerable interest during the
last two decades. Due to the rather high $K^+$-production threshold in
free nucleon-nucleon collisions ($T_{NN}{=}1.58$~GeV) and to the large
mean free path of $K^+$-mesons in nuclear matter, one hopes to extract
information about the intrinsic properties of the target nuclei and
in-medium properties of the kaons.  Obviously, in order to extract
this information, it is essential to determine the $K^+$-production
mechanisms.

Total $K^+$-production cross sections have been measured at the PNPI
synchrocyclotron for targets between Be and Pb and projectile
energies $T_p$ between 0.8 and 1.0~GeV~\cite{Koptev}, i.e., at beam
energies far below the free nucleon-nucleon threshold.  Inclusive
differential cross sections have been studied at
BEVALAC~\cite{Schnetzer}, SATURNE~\cite{Debowski} and
CELSIUS~\cite{Badala}.  Partial momentum spectra have been obtained at
different laboratory emission angles in the range
$10^{\circ}-90^{\circ}$, at projectile energies between 1.2 and
2.1~GeV.  Recently, full momentum spectra at forward emission angles
${<}12^{\circ}$ have been measured with the ANKE spectrometer of
COSY-J\"ulich \cite{Anke} at $T_p{=}1.0$ GeV. $K^+$ production has
also been studied at ITEP \cite{Akindinov,Buescher1}, where kaons with
fixed momenta and emission angle $10.5^{\circ}$ were identified at
projectile proton energies of 1.75 to 2.9~GeV.

The results were discussed in terms of different
models~\cite{Koptev,Cassing,roc,Sibirtsev,Paryev}, in particular of
single or two-step reactions, the latter with the creation of an
intermediate pion. It has been argued that the dependence of the
$K^+$-production cross section on the target mass $A$ is
sensitive on the mechanism dominantly contributing to kaon production.

In Sect.~\ref{sec:mass} it is studied whether a systematic analysis of
the $A$ dependence of the existing data allows, indeed, to draw
conclusions on the involved reaction mechanisms. In
Sect.~\ref{sec:spectra} it is shown that data taken at the same beam
energy $T_p$ but for different kaon momenta and emission angles can be
easily compared if plotted as a function of the four-momentum transfer
$t$ between the beam proton and the outgoing kaon.

\section{Target-mass dependence}
\label{sec:mass}
The first aim of our analysis is to evaluate the exponent $\alpha$ by
fitting the cross sections measured for different target nuclei with
an $A^\alpha$ dependence. We analyze available data on
$K^+$-production in $pA$ collisions at beam energies
$T_p{\le}2.9$~GeV. The data from Refs.~\cite{Badala,Buescher1} are not
taken into account here, since only a single target material was used
for these measurements. A similar analysis of the $A$ dependence for
meson production at high energies can be found in in
Refs.~\cite{Taylor,Johnson,Barton}.

The target-mass dependence for $K^+$-meson production in $pA$
collisions can be factorized in terms of the $A$ dependence of the
production process and that of the kaon propagation and distortion in
the nuclear medium.  Due to strangeness conservation, kaons with
relatively low momenta which are considered in this work, are not
absorbed in nuclear matter after their creation and can carry out the
information about the production mechanism. Also quasi-elastic
scattering of $K^+$-mesons in nuclei is substantially suppressed
leading to a large mean free path of $\lambda_{K^+}{\simeq}5$~fm.

The $A$ dependence for direct kaon production in collisions of the
beam proton with a single target nucleon is given by the inelasticity
of the $pA$ interaction. This also holds for the production of other
mesons, since the summation over all possible mesonic inelastic $pA$
reaction channels is proportional to the total inelastic $pA$ cross
section. The data~\cite{Afonasiev,Gachurin} on total inelastic
proton-nucleus cross sections can be well fitted by an $A^\alpha$
dependence with $\alpha{=}0.69{\pm}0.03$ at proton beam energies from
0.84 to 2~GeV. Modifications of the $A$ dependence can be due to the
intrinsic momenta of the participating nucleons, to Pauli blocking and
other nuclear effects, which essentially should not depend on the
target mass~\cite{Frullani}.  Therefore, one expects that the
$K^+$-production cross section is proportional to ${\simeq}A^{0.7}$ if
the kaons are dominantly produced via the direct production mechanism.

The $K^+$-mesons can also be produced in two-step mechanisms with
intermediate pion production in $pN_1{\to}\pi X$ reactions followed by
a $\pi N_2{\to}K^+ X$ process on a second target nucleon. Since two
nucleons are needed for the  kaon production, a stronger $A$
dependence for two-step production as compared with direct kaon
production is expected. Depending on the beam energy, $K^+$-production
may be due to both, the direct and two-step reaction mechanisms. At
low beam energies the two-step processes are energetically favorable
since the intrinsic nucleon motion can be utilized twice.

It has also been suggested that, in particular at deep subthreshold
energies, $K^+$-production is due to many-body interactions (like
e.g.\ the formation of clusters in the target nucleus) \cite{roc} or
is a reflection of a high degree of collectivity in the target nucleus
\cite{Anke}. We expect that such effects are proportional to the
number of target nucleons and, therefore, the cross section should
scale as ${\simeq}A^1$.

\begin{figure}[htb]
  \psfig{file=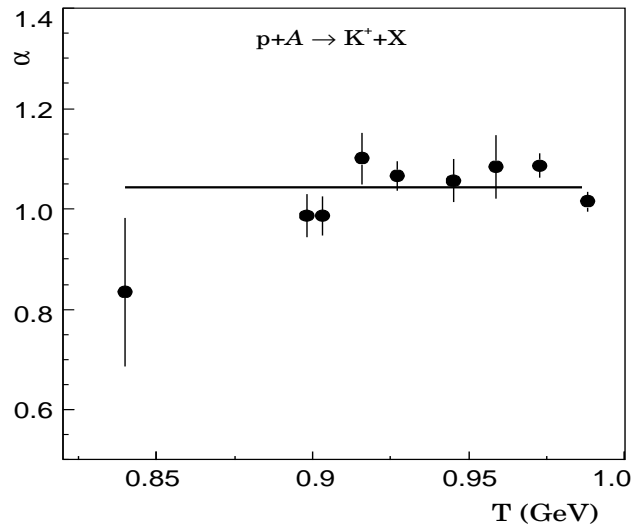,width=\figurewidth,height=\figureheight} 
  \vspace*{\figurespace}
  \caption{Dependence of the parameter $\alpha$ on the beam energy
    $T_p$ evaluated from the PNPI data~\protect\cite{Koptev} on total
    $K^+$-production cross sections. The solid line indicates the
    fitted average value $\alpha{=}1.04{\pm}0.01$.}
  \label{univ19a} 
\end{figure}

Total $K^+$-production cross sections $\sigma_{\mathrm{tot}}$ in
proton-nucleus collisions at $T_p{\leq}1.0$~GeV for $Be$, $C$, $Cu$,
$Sn$ and $Pb$ as target nuclei have been measured by Koptev et
al.~\cite{Koptev} at the Petersburg Nuclear Physics Institute (PNPI).
We recall that the threshold for $K^+$-production in free $pN$
collisions is 1.58 GeV and, thus, the PNPI measurements were devoted
to study deep-subthreshold strangeness-production mechanisms. At each
beam energy $\sigma_{\mathrm{tot}}$ was fitted by a function
$\sigma(A)=const  \cdot A^\alpha$.  Figure~\ref{univ19a} shows the
resulting parameters $\alpha$ as a function of $T_p$. In the full
energy range $0.842{\le}T_p{\le}0.990$ GeV, the data can well be
described by a constant value $\alpha{=}1.04{\pm}0.01$ as indicated by
the solid line and listed in the first line of Table~\ref{tab1}.  The
strong $A$ dependence of the total $K^+$-production cross section has
been interpreted \cite{Koptev,Cassing,Sibirtsev} as an indication for
the dominance of non-direct $K^+$-meson production in $pA$ collisions
at energies far below the free nucleon-nucleon threshold.

\begin{table}[htb]
  \caption{Values for $\alpha$ evaluated from the data on $K^+$-meson
    production in $pA$ collisions at various proton-beam energies
    $T_p$, kaon momenta $p_K$ and emission angles
    $\theta_K$.}
  \label{tab1}
  \begin{ruledtabular}
    \begin{tabular}{ccccr}
      $T_p$ (GeV) & $p_K$ (GeV/c) & $\theta_K$ ($^{\circ}$)&
      $\alpha$  & Ref.  \vspace{1mm} \\
      \colrule
      0.842--0.99 & $total$ & $total$ & $1.04{\pm}0.01$  & PNPI~\protect\cite{Koptev} \\
      2.1 & 0.35--0.75 & 15 & 0.56$\pm$0.05 & LBL~\protect\cite{Schnetzer} \\
      2.1 & 0.35--0.75 & 35 & 0.74$\pm$0.05 & LBL~\protect\cite{Schnetzer} \\
      2.1 & 0.35--0.75 & 60 & 0.84$\pm$0.08 & LBL~\protect\cite{Schnetzer} \\
      2.1 & 0.35--0.75 & 80 & 0.88$\pm$0.08 & LBL~\protect\cite{Schnetzer} \\
      1.2 & 0.5--0.7 & 40 & 0.69$\pm$0.3 & SATURNE~\protect\cite{Debowski} \\
      1.5 & 0.5--0.7 & 40 & 0.73$\pm$0.04 & SATURNE~\protect\cite{Debowski} \\
      1.7--2.91 & 1.28 & 10.5 & 0.54$\pm$0.02 & ITEP~\protect\cite{Akindinov} \\
      1.0 & 0.171--0.507$^{\ast}$ & 0--12 & 0.74$\pm$0.05 & COSY~\protect\cite{Anke}
    \end{tabular}
  \end{ruledtabular}
   {\small $^{\ast}$complete momentum spectrum covered}
\end{table}

\begin{figure}[htp]
  \psfig{file=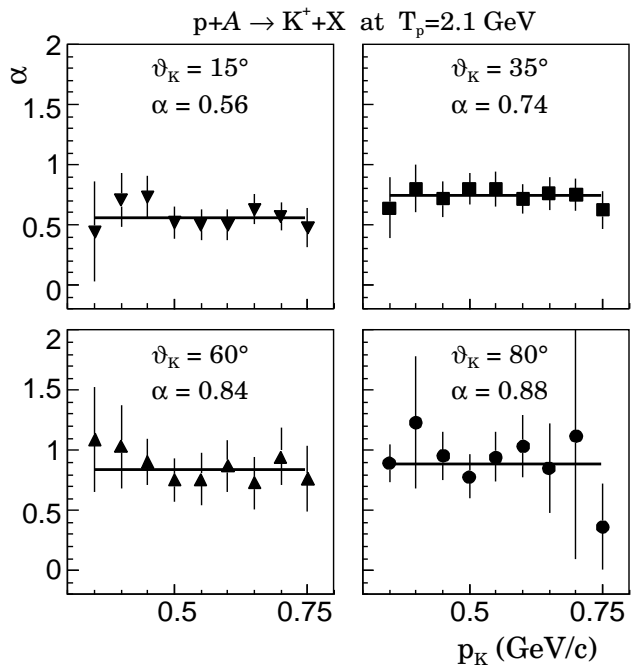,width=\figurewidth,height=\figureheightlong} 
  \vspace*{\figurespace}
  \caption{Exponent $\alpha$ as a function of the laboratory $K^+$-momentum
    evaluated from the LBL data~\protect\cite{Schnetzer} on
    differential $K^+$-production cross sections.  The solid lines
    show fits by a constant values $\alpha$.}
  \label{univ16a} 
\end{figure}

To obtain more detailed information about the strangeness-production
mechanisms, differential $K^+$-production cross sections were
subsequently measured by several groups.  Kaon production induced by
2.1 GeV protons (i.e.\ above the free $NN$ threshold) on $NaF$ and
$Pb$ targets has been studied by Schnetzer et al.~\cite{Schnetzer} at
the Lawrence Berkley Laboratory (LBL). The kaons were measured at
emission angles of $\theta_K{=}15{^\circ}, 35{^\circ}, 60{^\circ}$ and
$80{^\circ}$ and for momenta in the range 0.350--0.750 GeV/c.  We have
fitted the mass dependence of the double differential cross sections
by $d^2\sigma/dp\,d\Omega (A)=const \cdot
A^\alpha$ and show in Fig.~\ref{univ16a} the parameter $\alpha$ as a
function of the kaon momentum and for the different production angles.
Since no dependence of $\alpha$ on the kaon momenta is observed, we
fit $\alpha(p_K)$ by constant values. The results are shown in
Fig.~\ref{univ16a} by the solid lines and in Table~\ref{tab1}.

The mass dependence is weak ($\alpha{=}0.56{\pm}0.05$) for
$\theta_K{=}15^{\circ}$, and increases to larger angles
($\alpha{=}0.88{\pm}0.08$ for $\theta_K{=}80^{\circ}$). This might be
explained by the fact that the direct reaction mechanism dominates at
forward laboratory angles due to the Lorentz boost of this production
process.  It is also possible that the higher values of $\alpha$,
observed at larger angles, are related to a higher transparency of the
nucleus in case of particle production with high perpendicular
momenta. This phenomenon is well established in many processes and is
caused by the small size of the produced particles. In this case,
however, it starts at rather low momenta.  The dependence of $\alpha$
on the kaon momentum does not indicate a transition from the direct
production process to two- or multi-step $K^+$-production. In the full
momentum range, the LBL results are consistent with or close to the
expectation for direct $K^+$-meson production and the $A$ dependence
is  weaker than the one of the PNPI data~\cite{Koptev}.

\begin{figure}[htb]
  \psfig{file=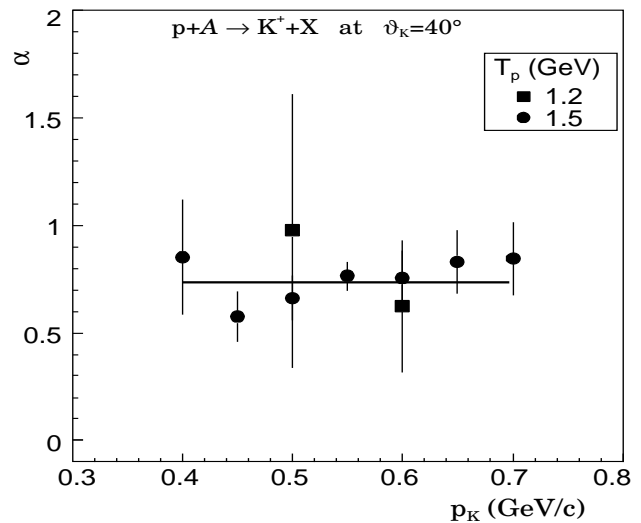,width=\figurewidth,height=\figureheight}
  \vspace*{\figurespace}
  \caption{$\alpha$ as a function of $p_K$ evaluated from the SATURNE
    data~\protect\cite{Debowski} on differential $K^+$-production
    cross sections. The different symbols corresponds to different
    beam energies $T_p$ as indicated in the figure. The solid line
    shows a fit by $\alpha{=}0.73{\pm}0.04$.}
  \label{univ17b}
\end{figure}

Double differential cross sections in $pC$ collisions at $T_p{=}1.2,
1.5$ and 2.5 GeV and in $pPb$ collisions at $T_p{=}1.2$ and 1.5 GeV at
a production angle $\theta_K{=}40^{\circ}$ have been measured by
Debowski et al.\ \cite{Debowski} at SATURNE.  The parameter $\alpha$
evaluated from the data at $T_p{=}1.2$ and 1.5 GeV is shown in
Fig.~\ref{univ17b} as a function of $p_K$. For both beam energies, the
results can be well fitted by a constant value
$\alpha{=}0.73{\pm}0.04$. In Table~\ref{tab1} the values for $\alpha$
are shown which were obtained from individual fits for 1.2 and 1.5
GeV. It can be seen that at 1.2 GeV the large error of $\alpha$ does
not allow any conclusions about the reaction mechanisms. At this
energy a dominance of two-step kaon production is expected
\cite{Debowski,Cassing,Sibirtsev}. At 1.5 GeV the relatively small
value of $\alpha{=}0.73{\pm}0.04$ is in line with the expected
dominance of one-step kaon production
\cite{Debowski,Cassing,Sibirtsev}.

\begin{figure}[htb]
  \psfig{file=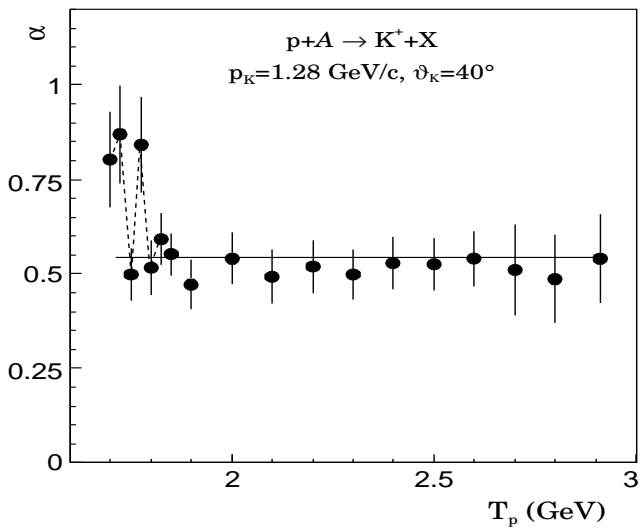,width=\figurewidth,height=\figureheight} 
  \vspace*{\figurespace}
  \caption{$\alpha$ as a function of the beam energy
    $T_p$ evaluated from the ITEP data~\protect\cite{Akindinov} on
    differential $K^+$-meson production cross sections for
    $p_K{=}1.28$~GeV/c and $\theta_K{=}10.5^{\circ}$.  The solid line
    shows a fit by a constant value $\alpha{=}0.54{\pm}0.02$. The
    dashed line only serves to guide the eye.}
  \label{univ18a} 
\end{figure}

Double differential cross sections for $K^+$-production with $Be$,
$Al$, $Cu$ and $Ta$ targets were measured by Akindinov et
al.~\cite{Akindinov} at the Institute for Theoretical and Experimental
Physics, Moscow (ITEP).  For $1.65{\le}T_p{\le}2.91$ GeV, kaons with
an emission angle $\theta_K{=}10.5^{\circ}$ and fixed momentum of
$p_K{=}1.28$~GeV/c were detected.  Below $T_p{\simeq}2.2$ GeV, such
high-momentum kaons cannot be produced in a free NN collision.  Thus
it is justified to assign these kaons to subthreshold particle
production.  We have fitted the $A^\alpha$ dependence and show
$\alpha$ as a function $T_p$ in Fig.~\ref{univ18a}.  The increase of
$\alpha$ at low beam energies is not understood so far. It maybe
either due to statistical fluctuations (indicated by the dashed line)
or reflect that below $T_p{\simeq}2.2$ GeV there is a transition from
direct to two-step $K^+$-production.  Within the experimental
uncertainties, the $A$ dependence of the ITEP data is compatible with
the LBL and SATURNE results.

The results on the $A$ dependence of kaon production are collected in
Table~\ref{tab1}.  The PNPI data on total $K^+$-production cross
sections at beam energies $0.842{\le}T_p{\le}0.990$~GeV indicate a
strong, ${\simeq}A^1$, dependence.  The data on differential cross
sections available for proton beam energies from 1.2 up to 2.9~GeV
yield an $A^\alpha$ dependence with $\alpha$ varying
between $0.54{\pm}0.02$ and $0.88{\pm}0.08$. Thus, for beam energies
$1.2{\le}T_p{\le}2.9$~GeV the average value of $\alpha{=}0.71{\pm}0.08$
is in reasonable agreement with the $A$ dependence expected from the
direct production mechanism.

The inconsistency between the PNPI~\cite{Koptev} data and the other
experiments~\cite{Schnetzer,Debowski,Akindinov} might be related to
the different beam energies. In fact, since the PNPI measurements were
performed at $T_p{<}1.0$~GeV, thus substantially below the kaon
production threshold in free space, two-step mechanisms or many body
effects should be more pronounced. To draw final conclusions one would
like to have data on differential cross sections at beam energies
close to 1.0~GeV.  The available data
\cite{Schnetzer,Debowski,Akindinov} indicate that the $A$ dependence
does not vary with the $K^+$-momentum. There is no indication for a
transition between direct strangeness production and the two-step
mechanism, which should cause a visible change in $\alpha$.

To clarify the situation, a measurement of $K^+$-meson production in
$pA$ collisions at a proton beam energy of 1.0~GeV was performed
recently with the ANKE spectrometer~\cite{Anke}.
Double differential  cross sections were measured at
emission angles of $0{\le}\theta_K{\le}12^{\circ}$ for $C$, $Cu$ and
$Au$ targets. The experiment was  designed for measurements
at forward angles, where the transition between the two-step mechanism
and direct $K^+$-production is expected to be most pronounced. In
fact, direct kaon production in $pN{\to}K^+X$ processes at far
subthreshold energies should entirely contribute at forward laboratory
angles.

\begin{figure}[htb]
  \psfig{file=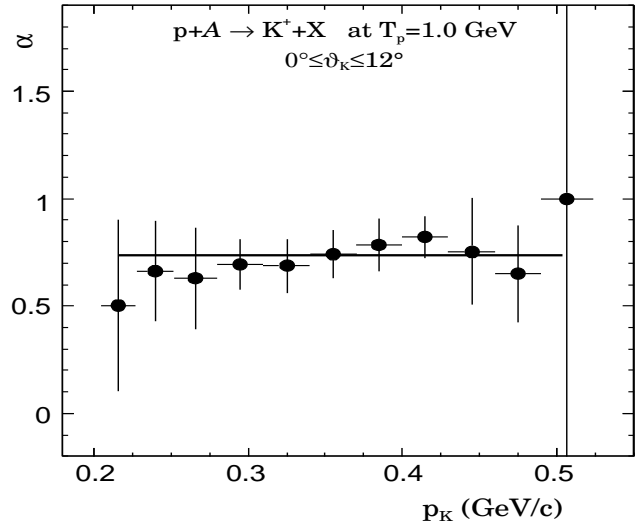,width=\figurewidth,height=\figureheight} 
  \vspace*{\figurespace}
  \caption{Dependence of  $\alpha$ on $p_K$ evaluated from the ANKE
    data~\protect\cite{Anke} at $T_p{=}1.0$~GeV and
    $0^{\circ}{\le}\theta_K{\le}12^{\circ}$. The solid line shows the
    fit by a constant value $\alpha{=}0.74{\pm}0.05$.}
  \label{mi4rb}
\end{figure}

Figure \ref{mi4rb} shows the exponent $\alpha$ as a function of the
laboratory $K^+$-momentum evaluated from the ANKE data.  The solid
line indicates a fit to the data with a constant value
$\alpha{=}0.74{\pm}0.05$.  A comparison with the values for $\alpha$
obtained from the total cross sections \cite{Koptev}
($\alpha{=}1.04{\pm}0.01$, see Table~\ref{tab1}) shows that the ANKE
data have a much weaker $A$ dependence. The only difference between
the PNPI and the ANKE measurements is that at ANKE the $K^+$-mesons
were detected inside the forward cone
$0^{\circ}{\le}\theta_K{\le}12^{\circ}$, whereas at PNPI
angular-integrated spectra were obtained.  Therefore, the very soft
$A$ dependence may be attributed to some special features of
$K^+$-production at forward angles. It may be speculated that at 1.0
GeV, $\alpha$ increases with increasing $K^+$-emission angle. A
similar behaviour has been deduced from the LBL data shown in Fig.\ 
\ref{univ16a}. However, these data were obtained at a significantly
higher beam energy where kaon production in single step reactions
should dominate. It would be interesting to check with new data from
ANKE whether at 2.1 GeV $\alpha$ is also significantly smaller at
angles around $0^{\circ}$.

We also find that the momentum-integrated cross section from ANKE for
angles $\theta_K{\le}12^{\circ}$ corresponds to roughly 10\% of the
total cross section, $\sigma_{\mathrm{tot}}=39$~nb for $pC$ collisions
at $T_p{=}0.990$~GeV \cite{Koptev}, whereas the covered solid angle
corresponds to less than 1\% of $4\pi$ \cite{ANKE-NIM}.  This
indicates a strong forward peaking of the produced kaons in the
laboratory system.

In contrast to previous studies at higher energies, the ANKE experiment
covers the full momentum range of the produced kaons at
1.0~GeV~\cite{Anke}. Figure \ref{mi4rb} illustrates that the data do
not show any dependence of $\alpha$ on $p_K$ for the complete
$K^+$-momentum spectrum.  Thus there is no indication for a change of
the $A$ dependence due to a possible transition from the direct to the
two-step production mechanism. We cannot provide any conclusive
explanation for this observation and for the low values of $\alpha$,
rather than that for the forward subthreshold $K^+$-production the
direct mechanism appears to be dominant.  This is in strong contrast
to results of model calculations \cite{Cassing,roc,Sibirtsev,Paryev}
where most of the kaons are produced in the two-step mechanism. It can
only be speculated here that two-step production combined with
rescattering effects of the produced kaons may result in a weaker $A$
dependence.

The results collected in Table~\ref{tab1} do not allow unambiguous
conclusions about the reacton mechanisms. In particular, there is a
strong disagreement between the 1.0~GeV data from ANKE~\cite{Anke}
measured under forward angles and the total cross sections from
PNPI~\cite{Koptev} obtained at the same beam energy. In order to
clarify the situation and to allow a comparison with the data on
differential cross sections obtained at higher
energies~\cite{Schnetzer,Debowski,Akindinov} it is necessary
to systematically measure kaon production at forward angles and at
higher energies.  These data would also help to reconstruct the
angular dependences of kaon production in $pA$ collisions at the
higher energies.

\section{Systematics of kaon spectra}
\label{sec:spectra}
At high energies data on hadron production in $pA$ collisions are
generally analyzed~\cite{Taylor,Johnson,Barton} in terms of the
Feynman variable $x_{\mathrm{F}}$ and the transverse momentum
$p_{\mathrm{t}}$ of the produced particle. The Feynman-scaling
variable is defined as
$x_{\mathrm{F}}{=}p_{\mathrm{l}}^*/p_{\mathrm{max}}^*$, where
$p_{\mathrm{l}}^*$ is the longitudinal momentum of the produced
particle in the center-of-mass system (CMS) of the incident proton and
the target nucleon, while $p_{\mathrm{max}}^*$ is the maximum CMS
hadron momentum  at a given beam energy.  However, one cannot
apply the commonly adopted $x_{\mathrm{F}}$ analysis of the $K^+$-data
at energies below the free $NN$ threshold, since a $pN$ CMS with the
nucleon at rest is kinematically not allowed.  One may select an
overall $pA$ CMS~\cite{Badala}, however then the joint analysis of
subthreshold kaon production and the data available above the $NN$
threshold becomes rather questionable.

In order to compare kaon spectra measured at different kinematical
conditions, i.e.\ proton beam energies $T_p$ as well as the kaon
momenta $p_K$ and emission angles $\theta_K$, we propose a more
natural kinematical variable given by the squared four-momentum
transfer $t$ between the produced kaon and the incident proton
\begin{equation}
  t{=}m_p^2{+}m_K^2{-}2\sqrt{(p_p^2{+}m_p^2)(p_K^2{+}m_K^2)}
  {+}2p_pp_K\cos\theta_K\ ,
\label{tdef}
\end{equation}
where $m_p$, $m_K$, $p_p$ and $p_K$ are the proton and kaon masses and
laboratory momenta, respectively, while $\theta_K$ denotes the
$K^+$-meson emission angle measured in the laboratory system
relatively to the direction of the proton beam. Since $t$ is Lorentz
invariant the analysis becomes independent of the choice of the
reference CMS, $pN$ or $pA$.
  
\begin{figure}[htb]
  \psfig{file=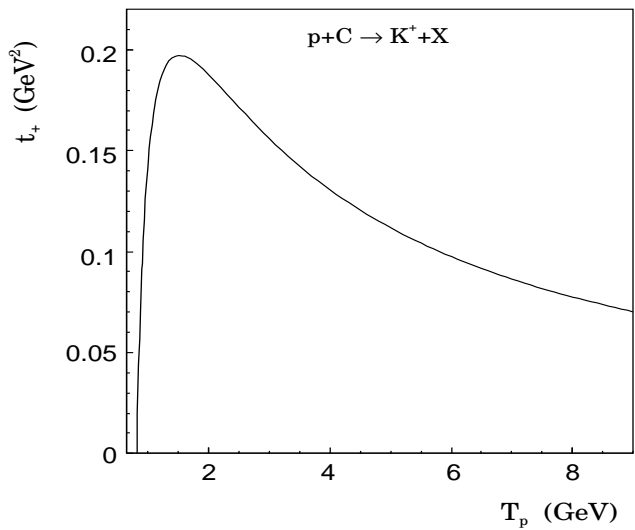,width=\figurewidth,height=\figureheight}
  \vspace*{\figurespace}
  \caption{Values for the maximum positive $t_+$ for
    $pA{\to}K^+X$ reactions as a function of the beam energy $T_p$.
    $t_+$ has been calculated for a minimum mass of the residual
    system $m_X{=}m_A{+}m_\Lambda$, neglecting binding energy of the
    $\Lambda$-hyperon in the target nucleus.}
  \label{fig:lola2} 
\end{figure}

A large four-momentum transfered from the incident proton to the
target (corresponding to large {\em negative} values of $t$) induces
excitation and disintegration of the target nucleus.  Reactions with
small $|t|$ are those where the produced $K^+$-mesons carry away a
substantial part of the momentum and energy of incident proton.  For a
$pA{\to}K^+X$ reaction the minimum and maximum values of $t$ 
follow from Eq.~\ref{tdef} as:
\begin{equation}
  t_{\pm}=m_p^2+m_K^2-2\sqrt{(q_p^2+m_p^2)(q_K^2+m_K^2)}\pm 2q_pq_K\ ,
  \label{tmax}
\end{equation}
where the sign of the last term corresponds to kaon production in
forward and backward direction relative to the proton-beam momentum
in the overall $pA$ CMS.  $q_p$ and $q_K$ are the proton and
$K^+$-meson three momenta  in the  $pA$ CMS,
\begin{eqnarray}
  q_p^2=\frac{(s-m_p^2-m_A^2)^2-4m_p^2m_A^2}{4s} \nonumber \\
  q_K^2=\frac{(s-m_K^2-m_X^2)^2-4m_K^2m_X^2}{4s}\ ,
  \label{moments}
\end{eqnarray} 
where $m_A$ denotes the target mass and $s$ the total invariant
energy of the beam proton with $T_p$ and the target
\begin{equation}
  s=m_p^2+m_A^2+2m_A(m_p+T_p)\ .
\end{equation}
In Eq.~\ref{moments} $m_X$ represents the invariant energy
of the residual final system with respect to the detected kaon.
The minimum and maximum values of $m_X$ are given as
\begin{eqnarray}
  m_X^{\mathrm{min}}=m_A+m_\Lambda \nonumber \\
  m_X^{\mathrm{max}}=\sqrt{s}-m_K\ ,
\end{eqnarray}
where $m_\Lambda$ is the $\Lambda$-hyperon mass. Note that by the
definition of $m_X^{\mathrm{min}}$ we neglect the $\Lambda$ binding
energy, i.e.\ formation of $\Lambda$-hypernuclei.
$m_X^{\mathrm{max}}$ corresponds to $K^+$-meson production with
transfer of all available collision energy to the invariant mass of
the residual system $X$ leading to $q_K{=}0$.  The maximum and minimum
values of $t$ are given by Eq.\ref{tmax} with minimum mass of the
residual system $m_X^{\mathrm{min}}$. The maximum positive squared
four-momentum transfer $t_+$ depends both on the beam energy $T_p$ as
well as the target mass $m_A$. Figure \ref{fig:lola2} shows the $T_p$
dependence of $t_+$ calculated for a carbon target and
$m_X^{\mathrm{min}}$. It is seen that $t_+$ can be positive and has
its maximum at small beam energies around $T_p{\simeq}1.5$~GeV.  It is
of specific interest to check whether processes with such positive
values of $t_+$ are experimentally accessible.

The $pA{\to}K^+X$ reaction amplitude ${\cal A}$ is a function of the
invariant collision energy $s$, the squared four momentum transfer $t$
and the squared invariant mass $m_X^2$ of the residual system.
Obviously, these Mandelstam variables can be expressed in terms of the
laboratory observables as the $K^+$-emission angle $\theta_K$ and
momentum $p_K$. Following Regge theory we factorize the $s$ dependence
of the reaction amplitude as
\begin{equation}
  {\cal A}(s,t,m_X^2) \propto f(t,m_X^2) \exp{[\gamma(t)\ln{(s/s_0)}]},
  \label{regge}
\end{equation}
where functions $f$ and $\gamma$ are given by the Regge pole with
quantum numbers of a strange baryon like a $\Lambda$ or $\Sigma$
hyperon and $s_0$ characterizes the domain of validity of the
Regge-pole theory. Assuming a linear $t$ dependence of the Regge
trajectory one can expand $\gamma(t){=}\gamma_0{+}\gamma_1t$, with
$\gamma_0$ and $\gamma_1$ being constant.

Within the beam-energy range $1.0{\le}T_p{\le}2.9$~GeV, $\ln{(s/s_0)}$
at $s_0{=}1$ GeV$^2$ varies between 5.14 and 5.37 and, therefore,
the reaction amplitude can be considered as a function of $t$ and
$m_X^2$ only.  In the following we investigate whether the $t$
dependence can be factorized out of both, the $s$ and $m_X^2$
dependences, for the $pA{\to}K^+X$ data. According to Eq.~\ref{regge}
we describe the invariant $K^+$-production cross section as
\begin{equation}
E \frac{d^3\sigma}{d^3p} = c_0\exp{[b_0t]},
  \label{eq:tdep1}
\end{equation}
with parameters $c_0$ and $b_0$ to be fitted to the data.
Finally, we intend to investigate the dependence of the slope
$b_0$ on the beam energy $T_p$ and the kaon emission angle 
$\theta_K$, which both are related to $s$ and  $m_X$. 

\begin{figure}[htb]
  \psfig{file=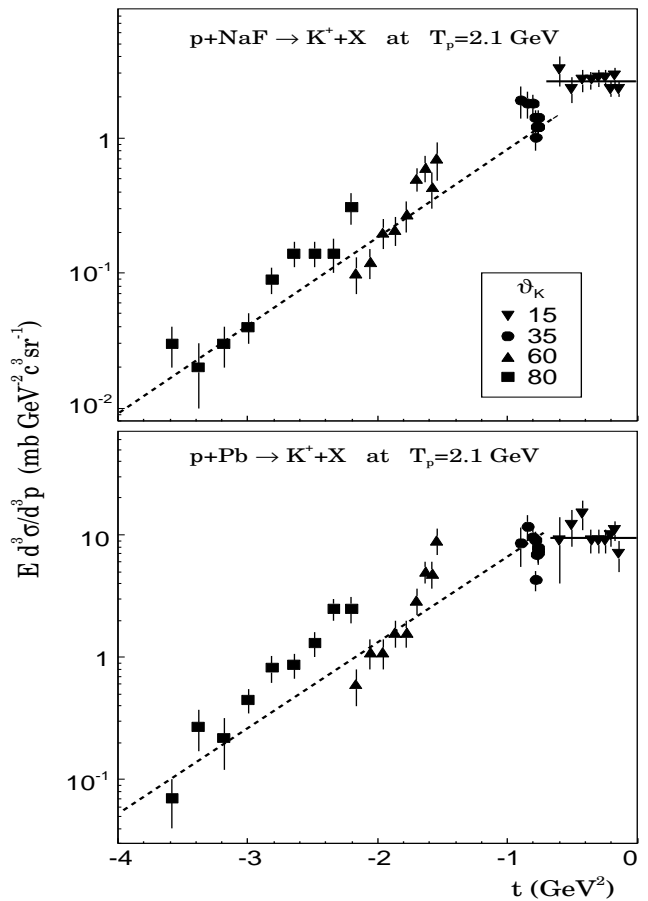,width=\figurewidth,height=\figureheighttwo}
  \vspace*{\figurespace}
  \caption{$t$ dependence of the invariant cross section for $NaF$
    (upper) and $Pb$ (lower) targets at $T_p{=}2.1$~GeV
    \protect\cite{Schnetzer}. The lines show a fit to the data by
    constant values (solid) and by Eq.\protect\ref{eq:tdep1} (dashed)
    with parameters listed in Table~\protect\ref{tabt}.}
  \label{mi1d} 
\end{figure}

Figure~\ref{mi1d} shows the invariant cross section
$E\,d^3\sigma{/}d^3p$ for $K^+$-production in $p(NaF)$ collisions at
$T_p{=}2.1$~GeV~\cite{Schnetzer}.  The measurements cover a $t$ range
from $-0.15$ to $-3.6$~GeV$^2$. For all measured angles and
$t{\le}{-}0.7$~GeV$^2$ the invariant cross section can be well fitted
by Eq.~\ref{eq:tdep1} with parameters $c_0$ and $b_0$ listed in
Table~\ref{tabt}.  At $-0.7{\le}t{\le}-0.15$ the invariant cross
section can be reasonably fitted by a constant value $c_1$, see
Fig.~\ref{mi1d} and Table~\ref{tabt}. The exponential dependence of
the invariant production cross section is not surprising and quite
typical for hadronic reactions.  However, the constant $t$ dependence
at low squared four-momentum transfer (small kaon emission angles)
seems to be unusual and needs further clarification.

\begin{table}[htb]
\caption{$t$ dependence  evaluated from the data on $K^+$-production
  in $pA$ collisions for different proton-beam energies $T_p$.
  $t_{\mathrm{min}}$ and $t_{\mathrm{max}}$ indicate the range of the
  squared four-momentum transfer measured in the individual
  experiments.  Parameters $c_0$ and $b_0$ were fitted with
  Eq.\protect\ref{eq:tdep1} to the data at large $|t|$, while $c_1$ was
  evaluated by fitting with a constant value at small $|t|$, see text.
  $c_0$ and $c_1$ are given in units of
  (mb$\cdot$GeV$^{-2}{\cdot}$c$^3{\cdot}$sr$^{-1}$).}
\label{tabt}
\begin{ruledtabular}
  \begin{tabular}{lccccccc}
    $T_p$ & $A$  & $t_{\mathrm{max}}$ & $t_{\mathrm{min}}$ &
    $c_0$  & $b_0$  & $c_1$  &  Ref.  \vspace{1mm} \\ 
    GeV & & GeV$^2$ & GeV$^2$ &  &
    GeV$^{-2}$ &  & \\
    \colrule
    2.1 & NaF & --3.58 & --0.15& 3.7$\pm$0.9 & 1.5$\pm$0.1
    & 2.6$\pm$0.1 & \protect\cite{Schnetzer} \\
    2.1 & Pb & --3.58 & --0.15& 32$\pm$8 & 1.6$\pm$0.1
    & 9.5$\pm$0.8 & \protect\cite{Schnetzer} \\
    1.2 & C & --0.41 & --0.48& (1.5$\pm$1.9) & 28$\pm$3
    & --- & \protect\cite{Debowski} \\
    & & & & $\times$10$^2$ & & \\
    1.5 & C & --0.59 & --0.64& (9$\pm$3) & 25.5$\pm$0.8
    & --- & \protect\cite{Debowski} \\
    & & & & $\times$10$^4$ & & \\
    2.5 & C & --1.22 & --1.65& (24$\pm$3) & 7.4$\pm$0.1
    & --- & \protect\cite{Debowski} \\
    & & & & $\times$10$^2$ & & \\
    1.2 & Pb & --0.41 & --0.45& (12$\pm$9) & 35$\pm$2
    & --- & \protect\cite{Debowski} \\
    & & & & $\times$10$^3$ & & \\
    1.5 & Pb & --0.58 & --0.7& (55$\pm$5) & 17.3$\pm$0.2
    & --- & \protect\cite{Debowski} \\
    & & & & $\times$10$^2$ & & \\
    1.2 & C & --1.10 & --1.25& (53$\pm$21) & 6.4$\pm$0.3
    & --- & \protect\cite{Badala} \\
  \end{tabular}
\end{ruledtabular}
\end{table}

The $t$ dependence of the invariant cross section for $K^+$-production
in $pPb$ collisions at $T_p{=}2.1$~GeV~\cite{Schnetzer} is shown in
the lower spectrum of Fig.~\ref{mi1d}. Again, the data show an overall
exponential scaling behaviour in $t$ and can be reasonably described
by Eq.~\ref{eq:tdep1} at $t{\le}{-}0.7$~GeV$^2$, while they are almost
constant at small $t$.  However, the $pPb$ data indicate some
systematic deviation from the exponential dependence at large angles
which can be attributed to function $f(t,m_X^2)$ in Eq.~\ref{regge}.
The slope $b_0$ obtained from the $pPb$ data is close to that fitted
to the $p(NaF)$ data, see Table~\ref{tabt}.

\begin{figure}[htb]
  \psfig{file=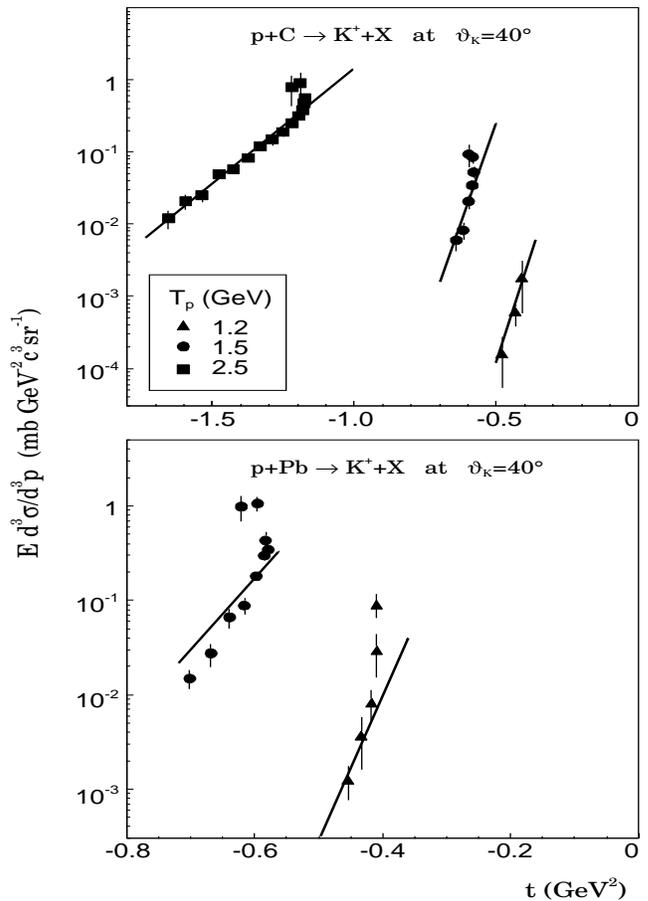,width=\figurewidth,height=\figureheighttwo}
  \vspace*{\figurespace}
  \caption{$t$-dependence of the invariant cross section for
    $pC{\to}K^+X$ reactions at $T_p{=}1.2, 1.5$ and 2.5~GeV (upper)
    and $pPb{\to}K^+X$ reactions at $T_p{=}1.2$ and 1.5 GeV (lower).
    The data were measured at $\theta_K{=}40^{\circ}$ and are taken
    from \protect\cite{Debowski}.  The lines show a fit to the data
    using Eq.\protect\ref{eq:tdep1} with parameters listed in
    Table~\protect\ref{tabt}.}
  \label{mi1f} 
\end{figure}

The $t$ dependence of the invariant cross section for $K^+$-meson
production in $pC$ collisions at $T_p{=}1.2, 1.5$ and 2.5~GeV and in
$pPb$ collisions at beam energies of 1.2 and 1.5~GeV \cite{Debowski}
is shown in Fig.~\ref{mi1f}.  It is found that the data clearly follow
an exponential dependence.  The fitted slopes $b_0$, see
Table~\ref{tabt}, substantially depend on the proton-beam energy. At
small $T_p$ the $t$ dependence is steeper. However, when comparing the
data with those from \cite{Schnetzer}, the different slopes cannot
completely be attributed to the variation of $T_p$, see first and
fifth line of Table~\ref{tabt}. The most probable explanation could
be essentially different excitations of the residual nuclei in these
cases. This problem clearly needs further inverstigation. At low $T_p$
the large values might be caused by the subthreshold production
mechanisms. With the presently available data it is not possible to
extract a systematic dependence of the parameters $c_0$ and $b_0$ on
$T_p$.

\begin{figure}[htb]
  \psfig{file=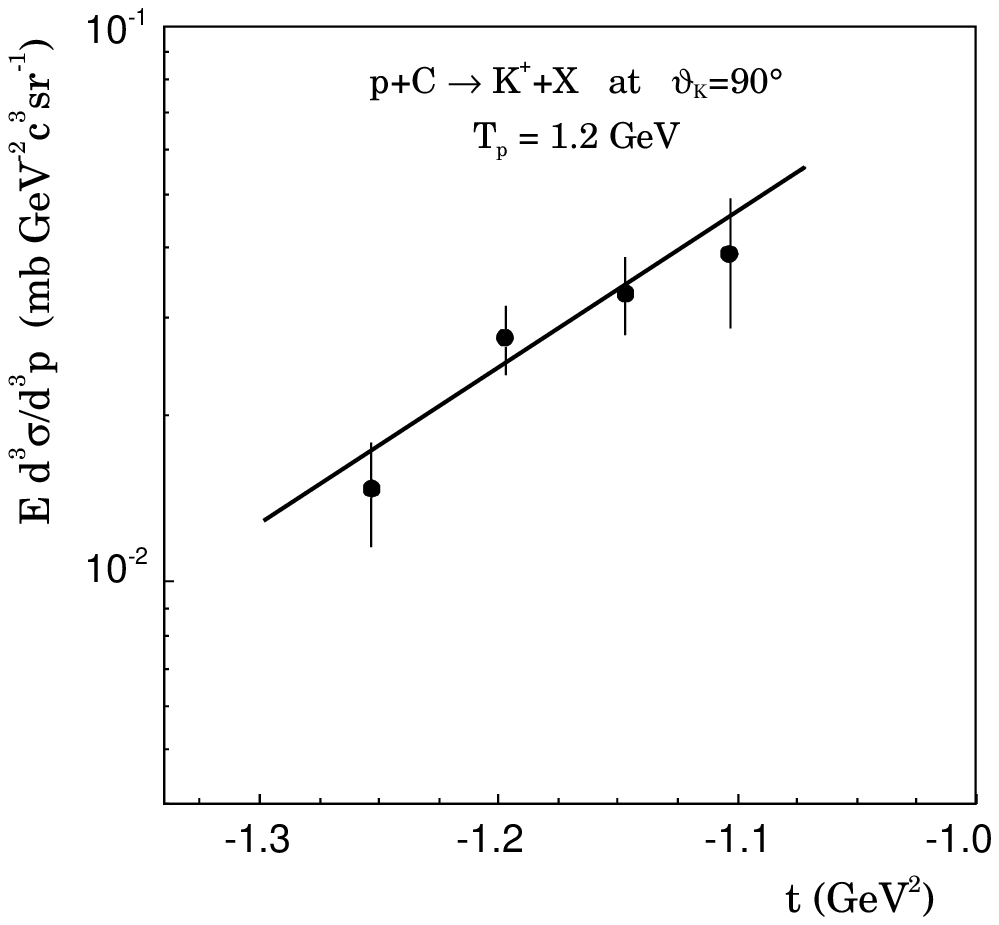,width=\figurewidth,height=\figureheight}
  \vspace*{\figurespace}
  \caption{$t$ dependence of the invariant $pC{\to}K^+X$
    cross section at $T_p{=}1.2$ and $\theta_K{=}90^{\circ}$
    \cite{Badala}. The line shows a fit to the data by
    Eq.\protect\ref{eq:tdep1} with  parameters listed in
    Table~\protect\ref{tabt}.}
  \label{mi1_bad} 
\end{figure}

Figure~\ref{mi1_bad} shows the $t$ dependence of the invariant
$K^+$-production cross section from $pC$ collisions at $T_p{=}1.2$~GeV
and $\theta_K{=}90^{\circ}$~\cite{Badala}.  Again, the data can be
well fitted by an exponential function with parameters given in
Table~\ref{tabt}.
 
\begin{figure}[htb]
  \psfig{file=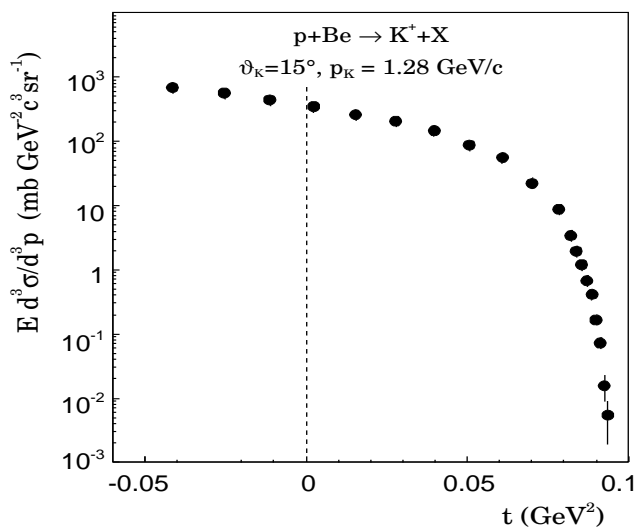,width=\figurewidth,height=\figureheight}
  \vspace*{\figurespace}
  \caption{$t$ dependence of the invariant $pBe{\to}K^+X$
    cross section~\cite{Akindinov} measured at $p_K{=}1.28$~GeV/c and
    $\theta_K{=}15^{\circ}$ at different beam energies.}
  \label{mi1k} 
\end{figure}

We also analyzed the data from Ref.~\cite{Akindinov}, where the $T_p$
dependence of $K^+$-production in $pA$ collisions was studied for
fixed kaon momentum $p_K{=}1.28$~GeV/c and production angle
$\theta_K{=}15^{\circ}$. The invariant cross section for $pBe$
collisions is shown in Fig.~\ref{mi1k} as a function of $t$. An
important feature of the data~\cite{Akindinov} is that they were
essentially collected at positive values of $t$. In principle, the
measurements probe the region around the maximum positive squared
momentum transfer $t_+$.

As we found before, the $t$ dependence can not be trivially factorized
from the $T_p$ dependence. For the data from Fig.~\ref{mi1k}, each
experimental point corresponds to a different proton-beam energy $T_p$
which, for fixed $p_K$ and $\theta_K$, can be calculated with
Eq.\ref{tdef}. Also $t_+$ is a function of $T_p$, and different for
each of the data points, which makes the analysis of the data
quite ambiguous.  Thus, with the data from Ref.~\cite{Akindinov} we
cannot verify the results for the data from Ref.~\cite{Schnetzer}
indicating a constant $t$ dependence of the invariant kaon production
cross section at $t{\simeq}0$.  We do not show the results
of~\cite{Akindinov} for the $Al$, $Cu$ and $Ta$ targets, which have
similar $t$ dependence as the data for $pBe$ collisions.

\begin{figure}[htb]
  \psfig{file=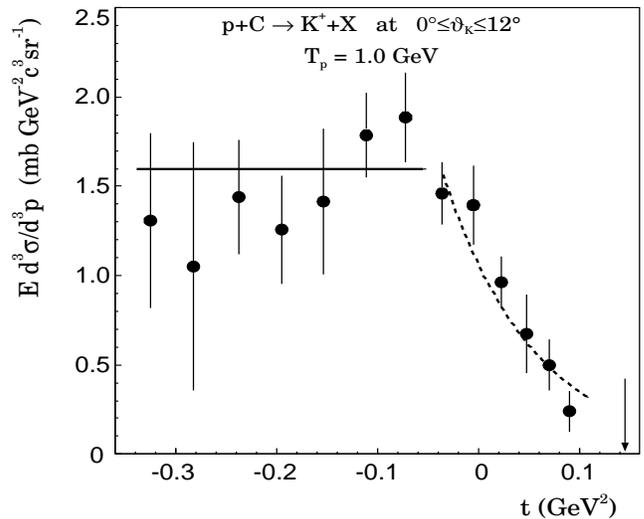,width=\figurewidth,height=\figureheight}
  \vspace*{\figurespace}
  \caption{The $t$ dependence of the invariant $pC{\to}K^+X$
    cross section~\cite{Anke} at beam energy $T_p{=}1.0$~GeV and kaon
    emission angles $0{\le}\theta_K{\le}12^{\circ}$.  The lines show
    the fit to the data by Eq.~\protect\ref{eq:tdep1} (dashed) and by
    a constant value (solid) with with the parameters from
    Eq.~\protect\ref{eq:tdep1}. The arrow shows the maximum possible
    value $t_+$, corresponding to kaons with the highest momenta. Note
    that the data are shown in a linear scale.}
  \label{mi1g} 
\end{figure}

Finally, the most recent results~\cite{Anke} on $K^+$-production in
$pC$ collisions at $T_p{=}1.0$~GeV and $0{\le}\theta_K{\le}12^{\circ}$
are shown in Fig.~\ref{mi1g}. It can be seen that the measurements
probe the region of $t{\simeq}0$, as well as close to $t_+$ from
Fig.~\ref{fig:lola2}.
The data confirm that at small $|t|$ the invariant $K^+$-production
cross section is almost independent of $t$.  Within the range
$-0.32{\le}t{\le}-0.07$~GeV$^2$ we fit the data by a constant value
$c_1{=}1.6{\pm}0.1\,
\mu$b$\cdot$GeV$^{-2}{\cdot}$c$^3{\cdot}$sr$^{-1}$.

For positive values of $t$ the dependence of the invariant
$K^+$-production cross section obviously changes and becomes
very steep in the range $-0.031{\le}t{\le}0.09$~GeV$^2$ where it can
be fitted by Eq.\ref{eq:tdep1} with
\begin{eqnarray}
  c_0=1.07\pm 0.08 \
  (\mu\mbox{b}\cdot\mbox{GeV}^{-2}{\cdot}
  \mbox{c}^3{\cdot}\mbox{sr}^{-1}), \nonumber \\
  b_0 =-11.1 \pm 1.7 \ (\mbox{GeV}^2)\ .
\label{para}
\end{eqnarray}
The strong decrease of the cross section towards $t_+$ is not
surprising since this region corresponds to the formation of
hypernuclei which characteristically is accompanied by very small
cross sections, see e.g.\ Ref.~\cite{COSY13}. 

Table \ref{tabt} shows that all available data taken at subthreshold
energies cover only very limited ranges of $t$, making systematical
analyses, like e.g.\ the extraction of the remaining $T_p$ and
$\theta_K$ dependences, impossible. Obviously, more data are needed at
different proton-beam energies $T_p$, as well as kaon angles different
from the previously measured ones. According to the reaction
kinematics shown by Eq.~\ref{tmax}, the extreme values $t_{\pm}$,
corresponding to large $t$ intervals, are accessible at forward
laboratory angles $\theta_K{\simeq}0^{\circ}$. In this respect the
ANKE spectrometer \cite{ANKE-NIM} is particularly useful since it
allows to measure kaon production at
$0^{\circ}{\le}\theta_K{\le}12^{\circ}$, and offers a wide
kaon-momentum coverage for beam energies in the range
$1.0{\le}T_p{\le}2.3$ GeV.

\section{Summary}
\label{sec:summary}
Our analysis of the target-mass dependence of $K^+$-meson production
in $pA$ collisions at $T_p{\le}2.9$~GeV show that, based on the
existing data \cite{Koptev,Schnetzer,Debowski,Akindinov,Anke}, it is
not possible to draw unambiguous conclusions about the underlying
reaction mechanisms. In particular, at $T_p{=}1.0$~GeV there is a
discrepancy between the PNPI data on total cross sections and those
from ANKE obtained under forward angles.  The data on differential
cross sections indicate that the parameter $\alpha$ only weakly
depends on the beam energy $T_p$ and kaon emission angle $\theta_K$
and is independent of the kaon momentum.  Surprisingly, for all these
data taken in the energy range $1.0{\le}T_p{\le}2.9$ GeV, the
target-mass dependence is close to the expectation for direct kaon
production.  We conclude that further $K^+$-momentum spectra should be
measured at forward angles and higher beam energies. The $A$
dependence of such data should be most sensitive to a transition from
one- to two-step reactions.  Furthermore, microscopical model
calculations should be performed which include rescattering effects of
the produced kaons in the nuclear medium.

The invariant $K^+$-production cross sections obtained under different
kinematical conditions show an overall exponential scaling behaviour
with the four-momentum transfer between the beam proton and the
produced kaon for $t{<}{-}0.05\ \mathrm{GeV}^2$. This indicates that
the Regge model is applicable at large negative values of $t$ even for
subthreshold kaon production down to $T_p{=}1.2$~GeV.  The dependence
on $t$ becomes steeper for small beam energies and kaon emission
angles $\theta_K$. It would be interesting to check by future
measurements whether for $t{<}{-}0.05\ \mathrm{GeV}^2$ the exponential
$t$ scaling is violated at $\theta_K{\simeq}0^{\circ}$. It can be
speculated that here the $t$ dependence becomes very strong.

At small negative values, $t{>}{-}0.05$, the invariant cross sections
are almost independent of $t$, if measured at the same beam energy
$T_p$. This indicates a transition from the Regge regime to
boson-exchange models. For $t{>}0$ the data show a strong falloff
towards the kinematical limit corresponding to the formation of
hypernuclei.

\acknowledgments This work profitted significantly from discussions
with members of the ANKE collaboration, in particular W. Cassing, M.
Hartmann and V. Hejny. Financial support by BMBF (WTZ grant
RUS-685-99) and RMS (grant FNP-125.03) is gratefully acknowledged by
one of the authors (V.K.). B.I. would like to thank J. Speth and A.
Sibirtsev for the hospitality at the IKP of the FZJ and the A.v.
Humbold foundation for financial support of his visit. His work is
supported in part by US CRDF (grant RP2-2247), INTAS (grant 2000-587)
and RFBR grant (00-02 17808).

\end{document}